# Augmenting Text Mining Approaches with Social Network Analysis to Understand the Complex Relationships among Users' Requests: a Case Study of the Android Operating System

Chan Won Lee, Sherlock A. Licorish, Bastin Tony Roy Savarimuthu and Stephen G. MacDonell
*Department of Information Science, University of Otago*
*PO Box 56, Dunedin 9016, New Zealand*
leech041@otago.ac.nz, sherlock.licorish@otago.ac.nz, tony.savarimuthu@otago.ac.nz,
stephen.macdonell@otago.ac.nz

## Abstract

*Text mining approaches are being used increasingly for business analytics. In particular, such approaches are now central to understanding users' feedback regarding systems delivered via online application distribution platforms such as Google Play. In such settings, large volumes of reviews of potentially numerous apps and systems means that it is infeasible to use manual mechanisms to extract insights and knowledge that could inform product improvement. In this context of identifying software system improvement options, text mining techniques are used to reveal the features that are mentioned most often as being in need of correction (e.g., GPS), and topics that are associated with features perceived as being defective (e.g., inaccuracy of GPS). Other approaches may supplement such techniques to provide further insights for online communities and solution providers. In this work we augment text mining approaches with social network analysis to demonstrate the utility of using multiple techniques. Our outcomes suggest that text mining approaches may indeed be supplemented with other methods to deliver a broader range of insights.*

## 1. INTRODUCTION

Understanding the needs and expectations of users has long played an important role in maximizing the usefulness and usability of delivered software systems. Gaining such an understanding therefore features prominently in contemporary business analytics, just as it has for decades. In their seminal work on usability [1], for instance, Gould and Lewis suggested that the number one determinant for producing successful systems is the consideration of users' needs as the main basis for system development. A subsequent study by Norman and Draper also supports the critical role that users play in developing and maintaining high quality systems [2]. Later approaches to delivering applications have thus looked to bridge the gaps between developers and users [3]. Such a strategy is also relevant to contemporary methods of systems development and deployment, which are driven by growing interest in virtual communities (e.g., Python development community[1]) and online application distribution platforms (OADPs, e.g., Apple's AppStore[2] and Google Play[3]). The burgeoning adoption of mobile devices means that users are now integrally involved in the development of apps[4]. In many cases users themselves are the architects of apps [4]. In addition, given the ubiquitous nature of virtual communities and social media, multiple communication channels that facilitate rapid communication between systems developers and target users are now available [5]. These channels enable (and even empower) users to describe their experiences while using apps, express satisfaction or dissatisfaction after using apps, rate apps and request new app features.

Thus, app owners are able to extract potentially meaningful insights from users' reviews with a view to improving the features they deliver to end-users. This process may not be straightforward, however, as large volumes of reviews can make this activity very challenging [6]. Thus, there is growing research effort aimed at aiding this process, and overcoming the need for extensive human involvement in the knowledge extraction process [7]. In particular, researchers have demonstrated that there is utility in various text mining approaches (e.g., those that reveal which features are said to be most in need of correction [8]) and modelling techniques (e.g., those that reveal the topics raised in relation to particular features [9]). Such approaches may be complemented by other techniques, to reveal not only what is most problematic or topical, but also how issues might be nested, or how issues concerning certain features may influence others. For instance, coupling between software modules A and B may mean that issues in module A may also affect module B. Revealing such relationships through the study of users' complaints and requests could inform maintenance strategies. We provide a demonstration of the use of complementary techniques in this work, investigating the Android OS[5] as a case study.

---

[1] https://www.python.org/community/
[2] itunes.apple.com/en/genre/ios/id36?mt=8
[3] play.google.com/store/apps?hl=en

[4] We use the term 'app' to describe a program or software product that is frequently delivered (especially to mobile devices) via an online platform.

In the next section (Section 2) we provide our study background, and we develop our hypotheses to address our objectives in Section 3. We then describe our study setting in Section 4. In Section 5 we present our results, and in Section 6 we discuss our findings and outline their implications. We then consider the threats to the validity of the work in Section 7, before providing concluding remarks in Section 8.

## 2. BACKGROUND

Understanding the content in online reviews has become integral to enhancing users' experiences, their satisfaction, and ultimately, maintaining a competitive advantage for the software provider [10]. However, gaining such understandings is challenging due to the labor-intensive nature of the insight extraction process from very large volumes of free text. In particular, and as mentioned above, the abundance of mobile devices and web-based communication channels means that it can be infeasible to manually extract insights from massive volumes of customer feedback [11]. While structured data such as rankings (or *likes*) and pricing details are easily analyzed, complexity is inherent, and rampant, in unstructured data or free text (e.g., customer reviews and opinions) [7]. Acknowledgment of this complexity has led to increasing interest in the potential utility of text mining.

Text mining is the extraction of information and knowledge from (very) large volumes of textual data using automated techniques [12]. Approaches typically provide contextual knowledge that may inform future project improvements, help individuals and organizations to avoid previous mistakes, and lead to improvements in customer service. Thus, text mining has the potential to make individuals and organizations aware of previously unknown facts and problem areas regarding a product [13].

Text mining can also take multiple forms. For instance, Iacob and Harrison [8] selected the top six most popular categories from Google's app store and extracted users' reviews automatically by using scraping tools (169 apps, 3279 reviews). The authors then manually explored the reviews in order to identify feature requests (totaling 763 requests). They observed 24 keywords in the feature requests. They then extracted all sentences in reviews that contained at least one of the 24 keywords, before applying a linguistic rule to identify specific requests. They next applied the Latent Dirichlet Allocation (LDA) model to identify the topics associated with particular features (e.g., *game* was associated with *update* and *levels*, where the authors inferred that users were interested in *games* having more *levels*) [8]. While Iacob and Harrison [8] used a linguistic rule and LDA, others have used part-of-speech tagging and sentiment analysis to achieve a similar goal, where nouns were extracted to study mentioned features and users' sentiments associated with such features [9].

One extremely large and successful community that has received considerable text mining attention is that of the Android OS. Studies have examined where bugs are most prevalent in the Android architecture [14], the quality of various types of bug reports [15], top user concerns and the attention they receive from the Android community [6], and the scale and severity of Android security threats as reported by its stakeholders [16]. With Android devices now leading mobile device sales [17], it is fitting that the text mining community would look to understand this platform and provide improvement advice. That said, we believe that the methods used for extracting knowledge from Android issue logs, and those used for understanding end-users' reviews logged on OADPs in general, could be usefully supplemented by approaches from the field of complex networks [18], and in particular, social network analysis [19], to discover useful patterns in this unstructured data. We outline two hypotheses to operationalize such an approach in the next section.

## 3. HYPOTHESES DEVELOPMENT

Previous studies have tended to employ text mining approaches in order to identify requests for specific feature enhancements [8, 9, 20]. However, due to the interconnected nature of software systems' components, dependencies between modules may mean that issues in one module could be interconnected with others, thus creating a web of potentially complex issues that may only be detected if the relationships among user requests or complaints are explored. In fact, those issues that are most frequently reported *in isolation* may not be as important to the community as those issues that have a wider effect through dependencies and other connections. Social network theoretical constructs, often used in social and behavioral sciences research to study the relationships between individuals and groups, have indeed noted that popularity (or count) as a measure may not adequately signify the importance or influence of an entity in a group, whereas closeness measures may be more indicative of these characteristics in network connectivity [19, 21, 22]. Thus, recommendations to address the most often requested features may not be optimal, especially in circumstances where features interact [6].

The opportunity to evaluate importance and influence is easily facilitated by the wealth of reviews that users provide, though the use of appropriate methods is essential for extracting this knowledge. Use of such approaches could help to decompose the complexity of users' expectations. In addition, such investigations have the potential to identify pointers for further in-depth investigations into online communities' challenges. With this in mind we outline our first hypothesis aimed at understanding the importance of popular feature requests, in terms of how they affect others.

---
[5] https://www.android.com/

*H1. Popular feature requests are not necessarily most important.*

Complex network theory notes that social and biological systems display interconnections that are unique, but not necessarily regular or random [18]. Although still developing, this theory has been applied to the study of software developers' dialogues and the artefacts they produce, with results confirming the usefulness of its application. For instance, studies have shown that analyzing developers' networks can improve bug prediction models [23]. Complex network theory has also been applied to the study of bug reporters' details to inform bug triaging procedures [24], and for understanding the effects of requirement dependency on early software integration bugs [25], providing promising outcomes. Given the way software components are often interconnected, issues in specific modules may influence others [6]. Users' complaints and requests may thus reflect this pattern, and therefore, in considering features in isolation, we may miss the complex webs of issues reported by users. We thus outline our second hypothesis.

*H2. Specific features in requests will strongly influence other feature requests.*

## 4. STUDY SETTING

We used the Android OS community as our case study 'organization'. Issues identified by the Android community are submitted to an issue tracker hosted by Google[6]. Among the data stored in the issue tracker are the following details: Issue ID, Type, Summary description, Stars (number of people following the issue), Open date, Reporter, Reporter Role, and OS Version. We extracted a snapshot of the issue tracker, comprising issues submitted between January 2008 and March 2014. Our particular snapshot comprised 21,547 issues, made up of defects and enhancement requests (e.g., "send message option is not available in speed dial recent list and it's annoying" and "provide an API for mounting a DocumentsProvider tree into a Linux directory"). These issues were imported into a Microsoft SQL database, and thereafter, we performed data cleaning by executing previously written scripts to remove all HTML tags and foreign characters [26], and particularly those in the summary description field, to avoid these confounding our analysis.

We next employed exploratory data analysis (EDA) techniques to investigate the data properties, conduct anomaly detection and select our research sample. Issues were labelled (by their submitters) as defects (15,750 issues), enhancements (5,354 issues) and others (438 issues). Given our goal we selected the 5,354 enhancement requests, as logged by 4,001 different reporters. Of the 5,354 enhancement requests, 577 were logged by members identifying themselves as developers, 328 were sourced from users, and 4,449 were labelled as anonymous. We examined the data of each request in our database to align these with the commercial releases of the Android OS. Its first release was in September 2008, although the first issue was logged in the issue tracker in January 2008. This suggests that a community was already actively engaged with the Android OS after the release of the first beta version in November 2007, with issues being reported just two months later. Given this level of active engagement, occurring even before the first official Android OS release, we partitioned the issues based on Android OS release date and major name change. So for instance, all of the issues logged from January 2008 (the date the first issue was logged) to February 2009 (the date of an Android release before a major name change) were labelled 'Early versions', reflecting the period occupied by Android OS releases 1.0 and 1.1 which were both without formal names. The subsequent partition comprised the period between Android OS version 1.1 and Cupcake (Android version 1.5), and so on.

Table 1 provides a brief summary of the numbers of enhancement requests logged between each of the major releases, from the very first release through to KitKat – Android version 4.4. From column three of Table 1 (Number of days between releases) it can be noted that the time taken between the delivery of most of the Android OS major releases (those involving a name change) was between 80 and 156 days, with two official releases (Gingerbread and Jellybean) falling outside this range. The fourth column of Table 1 (Total requests logged) shows that the number of enhancement requests reported increased somewhat as the Android OS progressed, with this rise being particularly evident when the mean requests reported per day for each release is considered (refer to the values in the fifth column for details). Over the six years of Android OS's existence, on average, 2.7 enhancement requests were logged every day (median = 2.6, Std Dev = 2.1).

We first employed natural language processing (NLP) techniques to study users' enhancement requests in evaluating our two hypotheses. NLP techniques are often used to explore and understand language usage within groups and societies. We employed multiple techniques from the NLP space in our analysis, including corpus linguistic part-of- speech tagging (POS) [27], computational linguistic n-gram analysis [28] and pointwise mutual information (PMI) measurement [29]. Thereafter, we used social network analysis (SNA) [22] to examine the way users' feature requests were interconnected. These procedures, and our associated reliability assessments, are now described in turn.

### 4.1. NLP Techniques

Prior research has established that noun terms in unstructured text reflect the main concepts in the subject of a clause [9]. From a POS perspective, nouns are indeed reflective of specific objects or things[7]. From a linguistic perspective, nouns often form the subjects and objects of clauses or verb phrases[8]. These and other understandings have been embedded as rules in NLP tools, including the Stanford parser which performs POS tagging [27]. We created a program that incorporated the Stanford API to enable us to extract noun phrases (features) from the enhancement requests, before counting the frequency of each noun as a unigram (e.g., if "SMS" appeared at least once in each of 20 enhancement requests our program would output SMS = 20). The ranking of words in this manner draws from computational linguistics, and is referred to as n-gram analysis. The n-gram is defined as a continuous sequence of n words (or characters) in length that is extracted from a larger stream of elements [28]. Adapting the PMI approach, which is used to measure the degree of statistical dependence between terms, we considered the syntactic relations between pairs of features (nouns) in each request by providing counts of these noun-pairs in the enhancement requests [29]. For example, if one issue read "the *Search* feature freezes when I try to access the *Map* app", and another "I am not sure that the *Search* feature optimizes the *Map* app", our output for the adapted PMI would be *Search-Map* = 2. To test our hypotheses the NLP outcomes were modelled using SNA techniques [22], as described next.

---

[6] code.google.com/p/android/issues/list

[7] www.merriam-webster.com/dictionary/noun

Table 1. Android OS enhancement requests over the major releases

| Version (Release) | Last release date | Number of days between releases | Total requests logged | Mean requests per day |
|---|---|---|---|---|
| Early versions (1.0, 1.1) | 09/02/2009 | 451 | 173* | 0.4 |
| Cupcake (1.5) | 30/04/2009 | 80 | 64 | 0.8 |
| Donut (1.6) | 15/09/2009 | 138 | 141 | 1.0 |
| Éclair (2.0, 2.01, 2.1) | 12/01/2010 | 119 | 327 | 2.8 |
| Froyo (2.2) | 20/05/2010 | 128 | 349 | 2.7 |
| Gingerbread (2.3, 2.37) | 09/02/2011 | 265 | 875 | 3.3 |
| Honeycomb (3.0, 3.1, 3.2) | 15/07/2011 | 156 | 372 | 2.4 |
| Ice Cream Sandwich (4.0, 4.03) | 16/12/2011 | 154 | 350 | 2.3 |
| Jellybean (4.1, 4.2, 4.3) | 24/07/2013 | 586 | 1,922 | 3.3 |
| KitKat (4.4) | 31/10/2013 | 99 | 781 | 7.9 |
|  |  | 2,176 | 5,354 | 2.7 |

\* Total number of requests logged between the first beta release on 16/11/2007 and Android version 1.1 released on 09/02/2009

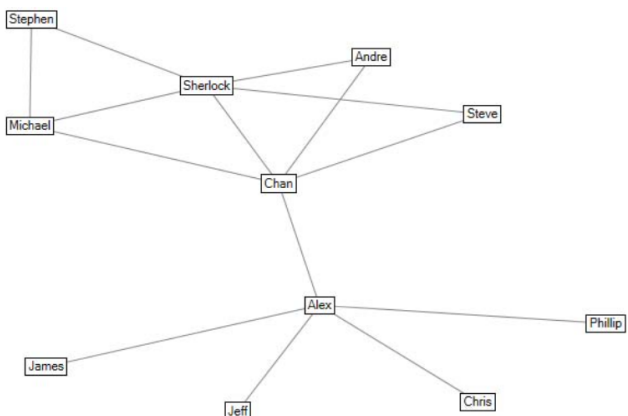

Figure 1. A simple social network

### 4.2. SNA Techniques
Social network analysis (SNA) is used to study the mapping of relationships between people, groups, organizations and other connected entities [19, 22]. In a sociogram (or SNA graph), people and groups are typically represented as vertices (nodes) while edges show relationships or information flows between the nodes (refer to Figure 1). Mathematical measures such as degree and closeness are often used to complement visual analysis, and can explain the influence of nodes in a network. In our work, we conducted social network analysis of the features (nouns) extracted in the reviews. We used three measures (corresponding to the variables considered in Section 3) to test our hypotheses, and these are considered in turn.

**4.2.1. Degree (popularity).** The *degree* of a vertex is a count of the unique edges that are connected to the vertex. This measure is also known as the "popularity measure" [19, 22]. The *degree* measure is represented formally as: degree ($v$) = E, where *degree* of vertex $v$ equals the number of edges $E$, that are connected to $v$. Figure 2 provides an illustration using a people network where the node *Andre* has a *degree* of 2 because he is only connected to *Sherlock* (degree=5) and *Chan* (degree=5). With the edges representing members' popularity, *Sherlock*, *Chan* and *Alex* are the most popular persons in the network; and *Phillip*, *Chris*, *Jeff* and *James* are the least popular with degree=1. We modelled undirected networks (edges with no arrows), as the order in which features are mentioned (i.e., directionality of the relationships) is not of significance in our study.

**4.2.2. Closeness (importance).** *Closeness* centrality measures the average shortest distance from one vertex to another. This measurement is computed to identify important nodes in a network. The higher the *closeness* centrality value, the shorter the distance from one vertex to other vertices (though in some cases the inverse is seen depending on the definition used [30]). For network optimization, if there is an item to be transferred around the network, the vertex with the largest *closeness* centrality measure is selected, given that this will be the fastest vertex to transfer the item to all other vertices. *Closeness* centrality is measured formally as: Closeness $(v) = (n-1) * CC(v)$, where $CC(v) = 1 / \sum_{y \in U} d(v,y)$, where $n$ is the number of vertices, $d(v,y)$ is the length of the shortest path between vertices $v$ and $y$ (the theoretic distance), and $U$ is the set of all vertices [31]. In Figure 1, *Chan* has the highest *closeness* centrality value, 0.67, indicating that this (Chan) vertex will transfer information the fastest to all other vertices on the network.

---
[8] www-01.sil.org/linguistics/GlossaryOfLinguisticTerms/WhatIsANoun.htm

**4.2.3. Clustering (influence).** The *clustering coefficient* often identifies the interconnectedness of a vertex with neighbor vertices, calculated by aggregating the total number of edges that are connected between a vertex's neighbors divided by the total number of possible connections between the vertex's neighbors [32]. *Clustering coefficient* is measured formally as: $Cv$ = *number of triangles connected to vertex v / number of triples centered around vertex v*, where two other vertices form a complete triangle with vertex *v* and a triple centered around vertex *v* is a set of two edges connected to vertex *v*. (If the degree of vertex *v* is 0 or 1, $Cv$=0). When all of a vertex's neighbors know each other the *clustering coefficient* is 1. This creates cliques or complete graphs. However, in some cases, not all neighbors are connected to each other. For example, in Figure 1, *Chan* has neighbors *Alex*, *Steve*, *Sherlock*, *Andre* and *Michael*. There are 10 triples around *Chan* (for example, *Chan-Alex-Steve*, *Chan-Alex-Andre*, *Chan-Alex-Sherlock*, and so on). However, *Chan* is only involved in three cliques/triangles (*Chan-Steve-Sherlock, Chan-Andre- Sherlock, Chan-Michael-Sherlock*), out of the 10 triples. The *clustering coefficient* is thus 3/10=0.3. A *clustering coefficient* closest to 1 indicates a strongly interconnected network.

We compute the three metrics just discussed using NodeXL[9] to test our hypotheses (employing degree to measure popularity, closeness to measure importance, and clustering coefficient to measure influence), on the network of features (nouns) obtained from enhancement requests. We performed reliability checks on the data used to generate the metrics introduced above, as described next.

**4.3. Reliability Assessments**
The first two authors of this work triangulated the NLP findings by manually coding a random sample of 50 outputs from the POS, n-gram and PMI analyses, to check that nouns and noun pairs were correctly classified (flagging each as true or false). We computed reliability measurements from these coding outputs using Holsti's coefficient of reliability [33] to evaluate our agreement. For our first reliability check (that nouns were correctly classified) we observed a 90% agreement, and the remaining 10% of codes were resolved by consensus. We observed 100% agreement for our second observation (where aspects were reported in conjunction). We report the results of our analysis next.

## 5. RESULTS
We report our results in this section. Firstly, we present results for popular feature requests in Section 5.1. We then provide results for important feature requests in Section 5.2, before finally considering results for feature requests that strongly influence others, in Section 5.3.

**5.1. Popular Feature Requests**

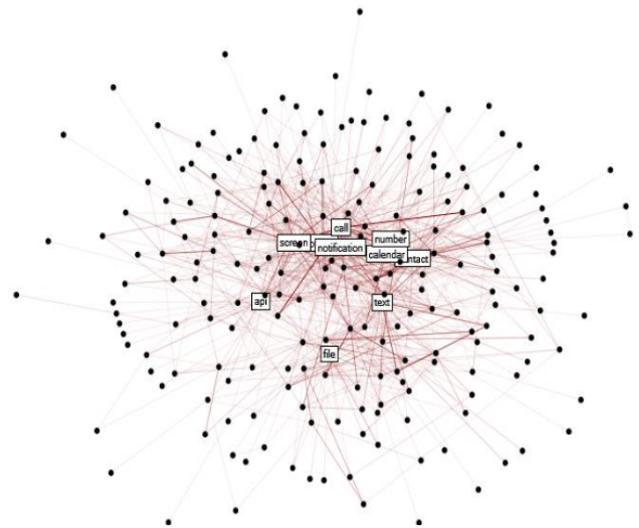

Figure 2. Network diagram for Android OS feature requests (where higher levels of concurrency are denoted by thicker edges)

We created a network diagram comprising 218 vertices (features) using the output from our PMI analysis (refer to Figure 2). Figure 2 shows that feature requests were highly interconnected. In Figure 2 we included the labels for the top 10 vertices, and edges were made thicker where there were higher levels of concurrency among reported features. We then ordered the vertices by their degree measures to assess features' popularity. Thereafter, we separated the vertices into three groups (two comprising 73 vertices and the third comprising 72 vertices) representing high popularity requests, medium popularity requests and low popularity requests. This approach was used previously to meaningfully examine differences across groupings [34]. We then selected the top 10 vertices (features) for each of the three groups based on their degrees for observation. These are presented in Table 2.

Table 2 shows that the most popular features (by count) in the network are screen (56), contact (55), call (50), notification (48), button (44), text (40), file (36), api (34), calendar (31) and number (30). This result reflects features that were requested most frequently for enhancements with others. In fact, considering the preliminary results from the n-gram analysis of the POS output which classified features on their own, we note a similar pattern for raw counts of requests [6], suggesting that those features that were mentioned most frequently for enhancements were also often mentioned with others in individual requests. We anticipated that, because these vertices were most frequently requested for enhancement, they would likely be connected to more features [31].

We thus examined the average degree of the overall network to see how the most popular vertices fared against the others (mean=8.4, median=5.0, Std Dev=10.2). When compared to the high vertices this measure was less than half (mean=19.2, median=15.0, Std Dev=11.5), nearly double that of the medium degree vertices (mean=4.8, median=5, Std Dev=1.4), and more than four times that of the low degree vertices (mean=1.5, median=1.0, Std Dev=0.6). We performed statistical testing to see if these differences were significant. We included the entire dataset in our tests, first assessing normality using the Shapiro-Wilk test, which revealed that our distributions (for high, medium and low degrees) violated normality (p<0.05). We then conducted a Kruskal-Wallis test, which confirmed that the differences in degree measures were significant (mean rank: high=182.5, medium=109.4 and low=37.6, p<0.01). These results suggest that the top enhancement requests were heavily skewed, and dominated the network by far. In addition, there were also significant differences between features grouped in the medium and low range (Mann-Whitney test result, p<0.01). In fact, looking at Table 2 it is noted that the top 10 features in the medium and low groups had degrees that were almost identical. Requests for enhancement in both groups were expressed less often with others (or alone [6]). We next considered the importance of these features in the network, and how they actually affected others.

---
[9] http://nodexl.codeplex.com/

Table 2. Degrees for high, medium and low vertices

| High | | Medium | | Low | |
|---|---|---|---|---|---|
| Vertex | Degree | Vertex | Degree | Vertex | Degree |
| screen | 56 | clock | 7 | dialog | 3 |
| contact | 55 | web | 7 | import | 3 |
| call | 50 | account | 7 | preference | 3 |
| notification | 48 | dock | 7 | swipe | 3 |
| button | 44 | gallery | 7 | version | 3 |
| text | 40 | word | 7 | root | 2 |
| file | 36 | dictionary | 7 | gradle | 2 |
| api | 34 | proxy | 7 | zoom | 2 |
| calendar | 31 | sim | 7 | launcher | 2 |
| number | 30 | studio | 7 | panel | 2 |

Table 3. Closeness for high, medium and low vertices

| High | | Medium | | Low | |
|---|---|---|---|---|---|
| Vertex | Close | Vertex | Close | Vertex | Close |
| screen | 0.53 | account | 0.39 | signature | 0.35 |
| notification | 0.52 | editor | 0.39 | gradle | 0.34 |
| call | 0.52 | address | 0.39 | vpn | 0.34 |
| contact | 0.51 | ringtone | 0.38 | homescreen | 0.34 |
| text | 0.50 | picture | 0.38 | authentication | 0.34 |
| button | 0.50 | adt | 0.38 | book | 0.34 |
| api | 0.49 | unlock | 0.38 | confirmation | 0.34 |
| data | 0.48 | web | 0.38 | facebook | 0.34 |
| calendar | 0.48 | sim | 0.38 | export | 0.34 |
| icon | 0.47 | gallery | 0.38 | family | 0.34 |

**5.2. Important Feature Requests**
We performed a similar analysis as that reported in the previous section in terms of separating requests based on importance. We examined the closeness centrality measure for vertices (features) to assess variances in how requests were connected in the network, as presented in Table 3. We outlined in Section 4.2.2 that the closest vertices in a network are often most integral to network connectivity. Table 3 shows that closeness for the top 10 high vertices range from 0.53 to 0.47. However, the differences in the ranges for the top 10 medium and low vertices are small (i.e., 0.01). We observe that eight of those features that were selected in the high degree range (see Table 2) were also most close (screen, notification, call, contact, text, button, api, calender), while there were four features that were selected in the medium degree range (Table 2) that appear in the medium closeness results in Table 3 (account, web, sim, gallery). That said, not all highly close features had high degrees.

In fact, of note is that seventeen of the highly close features were not listed among those with higher degrees when features were queried in our original dataset (comprising 218 vertices), suggesting that popularity indeed did not always confirm importance. We performed a similar Kruskal-Wallis test on the entire dataset as conducted in the previous analysis which confirmed statistically significant differences (p<0.01) in closeness measures across the three ranges (high, medium and low), with mean rank: high=182.0, medium=109.0, low=36.5. Follow up Mann-Whitney tests also confirm statistically significant differences across the closeness ranges (p<0.01 for all comparisons). Overall, these results suggest that degree counts on their own may not entirely reflect the most important requests.

**5.3. Feature Requests that Strongly Influence Others**
We adopted a similar partitioning to that used in the preceding subsections, and computed the clustering coefficient values for features. As noted in Section 4.2.3, when all of a vertex's neighbors know each other, the clustering coefficient is computed to be 1. Generally then, vertices with a clustering coefficient value of 1 are highly interconnected in their network segments. We note in Table 4 that those vertices occupying such highly connected segments were "birthday", "book", "caller", "emulator", "forwarding", "gesture", "gradle", "keystore", "launcher" and "microsoft". Of note is that "gradle" had a low degree in Table 2; however, like its higher closeness measure in Table 3, here we see that this feature was interconnected with many others (refer to Table 4). The "launcher" feature was similarly very connected with others but had a low degree (refer to Table 4). Security, library and frameworks and communication requests featured in the medium clustering coefficient category with values between 0.37 and 0.4, while those less connected had a clustering coefficient of between 0.17 and 0.20. We observe that "authentication" and "web" were close, and they had medium clustering coefficient values (0.40 and 0.38, respectively). Overall, the mean clustering coefficient of the graph we created was 0.37 (median=0.30, Std Dev=0.32). This suggests that 37% of all requests were connected to others. That said, a formal Kruskal-Wallis test confirmed statistically significant differences (p<0.01) in clustering coefficient scores across the three ranges (high, medium and low), with mean rank: high=182.5, medium=109.9, low=37.1. Follow up Mann-Whitney tests also confirm statistically significant differences across the groups (p<0.01 for all comparisons).

In fact, looking at the 218 vertices, we see that all of the features that had a high clustering coefficient in Table 4 had a degree less than or equal to 2. In addition, only "web", which had a medium degree in Table 2, had a similar placement in terms of clustering coefficient in Table 4. We also note here that degree measures did not appear to correlate with features' interconnections.

Table 4. High, medium and low clustering coefficients (CC)

| High | | Medium | | Low | |
|---|---|---|---|---|---|
| *Vertex* | *CC* | *Vertex* | *CC* | *Vertex* | *CC* |
| birthday | 1.0 | authentication | 0.40 | picture | 0.20 |
| book | 1.0 | encryption | 0.40 | sdk | 0.20 |
| caller | 1.0 | filter | 0.40 | server | 0.20 |
| emulator | 1.0 | gps | 0.40 | eclipse | 0.19 |
| forwarding | 1.0 | library | 0.40 | image | 0.19 |
| gesture | 1.0 | recognition | 0.40 | widget | 0.19 |
| gradle | 1.0 | webview | 0.40 | number | 0.18 |
| keystore | 1.0 | web | 0.38 | wifi | 0.18 |
| launcher | 1.0 | messaging | 0.37 | layout | 0.17 |
| microsoft | 1.0 | background | 0.37 | button | 0.17 |

Table 5. Main feature requests clusters

| Rank | Size | Description: Examples |
|---|---|---|
| 1 | 67 | *Sound/Music*: mp3, track, volume, ring. *Hardware*: slider, pc, usb, power, headset. *Screen*: wallpaper, icon, brightness, touch. |
| 2 | 63 | *Communication*: email, voicemail, contact, sms. *Calendar*: birthday, date, schedule. *Notification*: ringtone, vibration. |
| 3 | 59 | *Language*: spanish, hindi, dictionary, documentation. *Built-in function*: library, mouse, flash, emulator. |
| 4 | 16 | *Web/Security*: webview, authentication, proxy. *User actions*: navigation, sensor, status. |
| 5 | 13 | *Call*: caller, dialer, log, dialog, number. *Display*: image, panel. |

We went one step further and used the Clauset-Newman-Moore algorithm to partition the network into clusters that demonstrated higher numbers of interconnections than others [32]. We provide these groupings in Table 5, ranked based on the size of the group. Here it is noted that the highest ranked group inside the network contained 67 vertices (features). Examining the actual requests we observed that the categories related to sound/music, hardware and screen (refer to the third column). On the backdrop that Clauset-Newman-Moore clusters vertices that are densely connected [32], requests grouped together thus have very strong relationships. In fact, in considering our case, it is plausible that requests for enhancements to sound, hardware and screen features are related. We observed that the second cluster contained 63 vertices, with requests related to communication, calendar and notification features (refer to Table 5). The third cluster contained 59 vertices, with requests for enhancements to language features and built-in functions. The fourth cluster contained 16 vertices that related to web/security and user actions, while the fifth cluster contained 13 feature requests that were related to call and display features (refer to Table 5).

We observe here users' feature requests that are derived through the use of NLP and SNA techniques that require less human involvement to perform data processing (than say [8]). However, the stages of their revelation could be much more informative than the commonly used LDA technique [20]. For instance, LDA assigns topics to documents using a probabilistic distribution algorithm where documents are modelled as a mixture of topics [35]. If all reviews are entered as one input file to the LDA algorithm, then the outcome of this algorithm would likely be different features and their associations with many topics (as evident in all the reviews combined). These topics may not necessarily be the features they co-occur with most frequently. In fact, even if each review is mined as a document and submitted to the LDA algorithm, specific words may still have much higher associations with certain features than other features, making it difficult to identify feature-feature co-occurrences.

## 6. DISCUSSION AND IMPLICATIONS

*H1. Popular feature requests are not necessarily most important.* We observed that feature requests were highly interconnected, suggesting that issues that affected community users using software were potentially similarly interconnected as software modules usually are. In addition, specific features were requested for enhancement more often than others. However, these features were not always highly connected to others, so remedial work undertaken on these features may be less complex than on features that affected many others. In fact, popular features may not necessarily be the most important, in terms of being highly connected to many others. Our evidence confirms the long established observation in SNA that popularity (or count) as a measure may not adequately signify importance [19, 22]. We must therefore accept H1.

Given our findings, developers may consider numerous factors when assessing the importance of users' feature requests in OADPs, and which features should be prioritized for work. Among other considerations, a strategy for effecting revisions may consider the cost of human effort involved in the rework, the return to be had in investing such effort, other competing priorities, how many users are affected by the issue (i.e., number of complaints), which group of stakeholders is affected the most by the issue, and so on. In considering the number of complaints, perhaps a strategy that considers how such complaints affect others would also be useful. For instance, the most requested features for enhancements in Table 2 (see first column) had less impact on many other features (see first column of Table 4). In contrast, those features in column one of Table 4 always affected others. Given this finding, with competing priorities, perhaps a strategy aimed at fixing those features in Table 2 would not be the most cost-effective, considering that those in Table 4 (and Table 3) are affecting many others.

*H2. Specific features in requests will strongly influence other feature requests.* Some requests tend to co-occur, with this co-occurrence being evident across application layers (Android OS modules are generally arranged in a layered structure, with the Application Layer at the top, followed by the Application Framework layer, Libraries and the Linux kernel [17]). For instance, language features were clustered with library features in Table 5. Our evidence suggests that possible interconnections in software components mean that issues in specific modules influence others. Users' feature requests thus reflect this pattern, and therefore, developers are encouraged to not consider these requests in isolation, which may result in them missing complex webs of issues that are often reported by users. Given our findings we must accept H2.

Our evidence suggests that techniques that help developers to reveal the interconnections in users' complaints could be relevant in informing software revision/improvement strategies. We observe here an exposure of users' feature requests that was derived through the use of NLP and SNA techniques that require less human involvement to perform data processing (than say [8]). SNA has also demonstrated utility beyond its previous use for understanding communication [23]. In fact, the outcomes from the application of the combined techniques used in this work is comparable to the well-known LDA technique [20] (refer to Section 5.3 for details). While this is just one demonstration of the utility of other approaches for revealing the complex connections among features, we believe that supplementing text mining techniques with other tools has the potential to reveal useful patterns in unstructured data, which could be enlightening to the OADP community.

# 7. THREATS TO VALIDITY

Although the Android issue tracker is publicly hosted, and so is likely to capture a wide range of the community's concerns [14], issues may also be informally communicated and addressed within the development teams at Google. Similarly, unreported issues are not captured by our analysis. We also accept that there is a possibility that we could have missed misspelt features. However, our reliability assessment measure revealed excellent agreement between coders [33], suggesting that our findings benefitted from accuracy, precision and objectivity.

# 8. CONCLUSION

Responding to users' concerns helps to ensure that systems address users' needs. We noted that, with recent approaches allowing the distribution of software systems online to a wide user-base, responding to users' needs and expectations could be potentially challenging. Thus, text mining and automated approaches are used increasingly to reduce the burden associated with humans exploring large volumes of users' reviews. Most often, such techniques reveal the issues that are mentioned most often for correction, and the topics that are associated with particular features. In this work we anticipated that SNA techniques might provide additional insights in terms of how users' reviews are interconnected. This knowledge, we proposed, could be useful in identifying critical issues and informing systems revision strategies. We thus used social network theoretical constructs as a basis for studying the importance of popular feature requests and how users' requests influence others. We found that popular features were not necessarily the most important, in terms of being highly connected to many others, and that some requests co-occurred. We observed users' feature requests patterns of interconnectedness, suggesting that outcomes from investigations examining such trends could be useful for informing revision strategies. Overall then, we believe that text mining should be supplemented with other approaches to provide advanced analytics.